\documentstyle[prl,aps,epsfig]{revtex}         
\begin{document}
\draft               
\twocolumn[\hsize\textwidth\columnwidth\hsize\csname @twocolumnfalse\endcsname

\title{Coherent switching of semiconductor resonator solitons }
\author{V. B. Taranenko, F.-J. Ahlers, K. Pierz}
\address{Physikalisch-Technische Bundesanstalt 38116 Braunschweig, Germany}
\maketitle
\begin{abstract}
We demonstrate switching on and off of spatial solitons in a
semiconductor microresonator by injection of light coherent with
the background illumination. Evidence results that the formation
of the solitons and their switching does not involve thermal
processes.
\end{abstract}
 \vskip1pc ]
Spatial solitons in semiconductor resonators have recently been
considered for applications as mobile binary information carriers
for certain optical parallel information processing tasks
\cite{tag:1,tag:2}. We have recently found that such dark and
bright spatial solitons can exist in nonlinear passive
semiconductor resonators \cite{tag:3}. We demonstrated in
\cite{tag:4} that bright solitons can be switched on (and also
temporarily be switched off) by addressing with a light beam
incoherent with the background light field, which sustains the
solitons. In this case a light-induced increase of temperature
\cite{tag:5} is involved in both switching on and off which
limits the switching speed to typically 1 $\mu$s.

An alternative mechanism of switching such solitons on or off is
by changing the carrier density in the resonator at the location
of the soliton. This can be effected by addressing the position
of the soliton (to be created or to be switched off) with a
sharply focused light beam coherent with the background light.
The latter's intensity is to be within the coexistence range of
soliton and low transmission resonator state. If the address beam
is in phase with the background light the intensity at the
soliton location will be increased when admitting the address
beam. This increases the carrier density locally and therefore
can switch a soliton on. Vice versa, an existing soliton can be
switched off when the address beam is in counterphase to the
background light. By destructive interference the light intensity
at the location of the soliton can decrease below the existence
range of solitons so that a soliton may be switched off.

In this coherent scheme primarily the changing of the local
carrier density causes the switching, whereas in the incoherent
scheme reported in \cite{tag:4} the increasing of the temperature
causes the switching. Nonthermal incoherent switching can also be
done - e.g. by application of local optical pumping \cite{tag:6}
pulses which increase the carrier density. Nonetheless the
coherent scheme is the only one permitting on- and off-switching
in a straightforward manner \cite{tag:2,tag:7,tag:8,tag:9}. We
conducted therefore an experiment to demonstrate this mechanism.\\

Fig.~1 shows the optical arrangement used for the observations.
Light of single frequency Ti:Al$_2$0$_3$-laser at 850 nm
wavelength illuminates an area on the resonator sample of $\sim$50
$\mu$m diameter. Part of the light is split away from the main
beam and then superimposed with the main beam by a Mach-Zehnder
interferometer arrangement, to serve as the address beam. In the
address path of the interferometer the light is expanded so that
the focus of the address beam is narrow ($\sim$ 8 $\mu$m). An
electro-optic modulator switches the address beam on for a short
time ($\sim$ 40 ns). One of the interferometer mirrors can be
moved by a piezo-electric element to control the phase difference
between the background light and the address light.

\begin{figure}[htbf] \epsfxsize=82mm
\centerline{\epsfbox{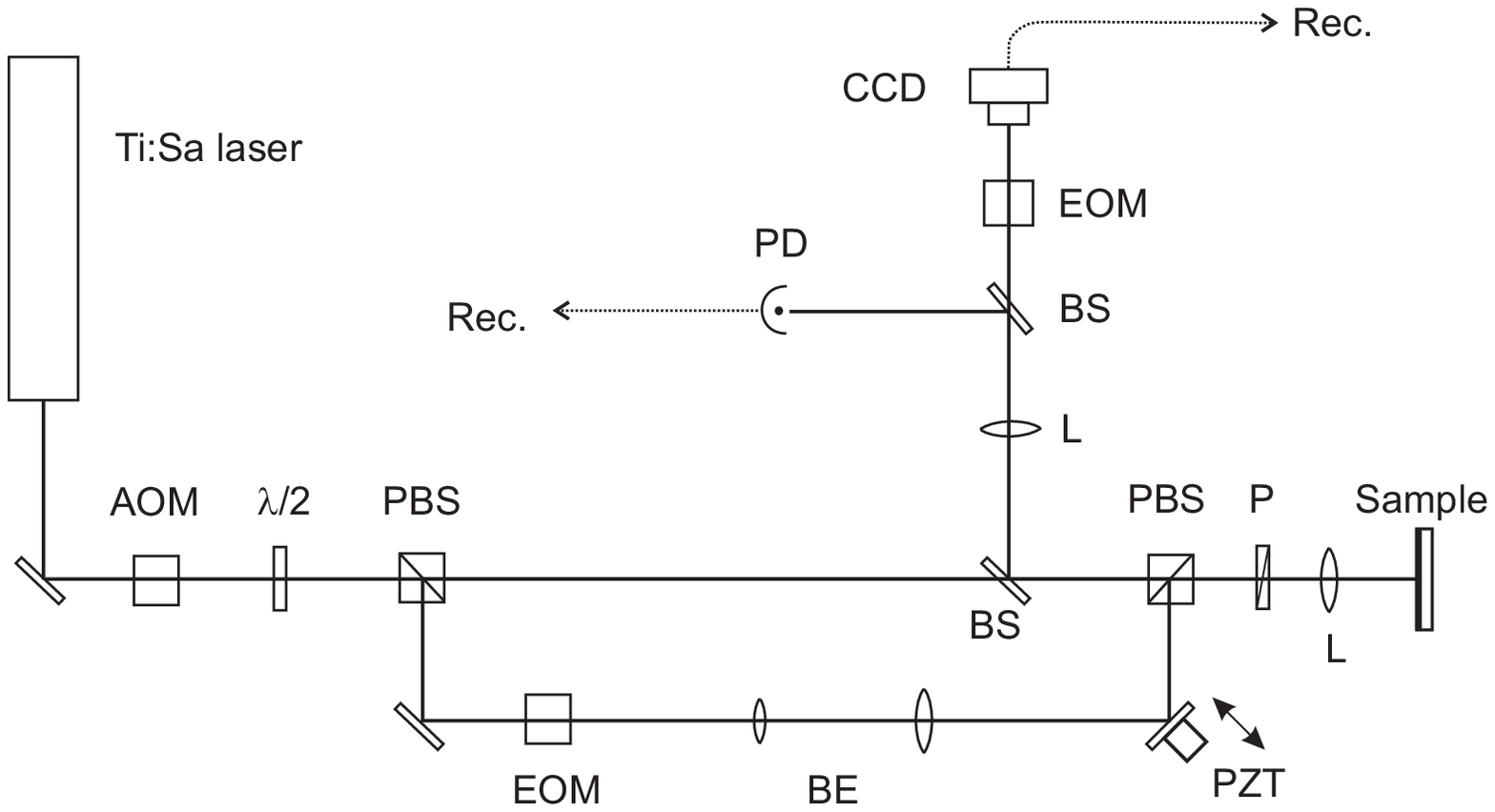}} \vspace{0.7cm}
\caption{Experimental setup.\\ AOM: acousto-optic modulator,
$\lambda$/2: halfwave plate, PBS: polarizing beam splitters, EOM:
electro-optical amplitude modulators, BE: beam expander, PZT:
piezo-electric transducer, L: lenses, BS: beam splitters, PD:
photodiode, P: polarizer.\\ Polarizer P combines coherently the
injection and background fields.}
\end{figure}

In order to avoid long-term thermal drifts the measurements are
carried out within a few $\mu$s. The laser light is admitted
through an acousto-optic modulator for durations of a few $\mu$s
and repeated every ms. For observation a CCD-camera combined with
an electro-optic modulator serves to take snapshots on a
nanosecond timescale. The modulator opens with a variable delay
after the start of the illumination so that snapshots can be
taken at different times during the illumination. The dynamics is
followed more directly locally by imaging a fast photodiode onto
the center of one soliton on the resonator sample. The light
intensity at the soliton location is then recorded as a function
of time.

The semiconductor sample was a passive resonator containing 12
GaAs quantum wells cladded by two Bragg reflector stacks
consisting of 23 and 17 pairs of AlGaAs/AlAs quarter-wave layers,
respectively. The Bragg mirrors were designed to be
non-absorptive above 810 nm. The resonator itself was a 3/2
$\lambda$ resonator of the 'cos' type and the width of the GaAs
quantum wells was 15 nm. Suitable spacers between the QW material
and the Bragg stacks place the resonance wavelength of this
resonator in the vicinity of the band gap at 850 nm. The radial
thickness variation of the layers from the center to the edge of
the circular 2 inch wafer causes a radial variation of the
resonance wavelength of only 7 nm. This very good large-scale
homogeneity allows to maintain the relative detuning of laser
light to resonance center over large distances on the wafer. This
kind of sample should therefore be well suited for experiments on
motion of spatial solitons. The resonance wavelength everywhere
on the wafer is close to (below) the band edge wavelength. Thus
the nonlinearity is strongly absorptive.\\

Working at $\lambda$ = 850 nm, bright spatial solitons are readily
observable (Fig.~2); they form spontaneously or with addressing
pulses. Figs.~3 and 4 show the results of the switching
experiments. The dotted lines show the incident intensity and the
solid line the intensity reflected from the sample. Note that the
observation is in reflection such that a bright soliton (high
transmission) shows up as a reduction in reflected intensity.

\begin{figure}[htbf]
\epsfxsize=70mm \centerline{\epsfbox{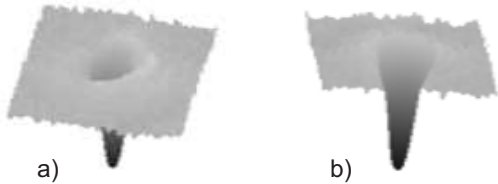}} \vspace{0.7cm}
\caption{Bright soliton (dark in reflection). 3D representation
of reflectivity: a) view from above, b) view from below.}
\end{figure}

Fig.~3a shows the switch-on. The background intensity is chosen
slightly below the spontaneous switching threshold. With the
application of the in-phase address pulse the reflected light
intensity increases (constructive interference) and the soliton
is switched on (evidenced by the reduced reflectivity). Evidently
the soliton remains switched on until the background light is
reduced to the lower limit of the existence range of the soliton.

Fig.~3b shows a "failed switch" in which the intensity of the
address beam is too low to switch. Equally to Fig.~3a the
in-phase address pulse shows up as increased reflected light
intensity, however, the total intensity at the address position
does not reach the upper limit of the soliton existence range so
that no soliton forms.

Fig.~3c shows the switching off of a soliton. The background
light intensity is chosen above the coexistence range of soliton
and low transmission resonator state, so that a soliton forms
spontaneously. The address beam is then applied in counter-phase
to the background field (evident by a reduction of reflected
light due to the destructive interference between background- and
address light), the soliton disappears, and the resonator returns
to the low transmission state. Fig.~3d shows a "failed
switch-off" analogously to Fig.~3b.

\begin{figure}[htbf]
\epsfxsize=85mm \centerline{\epsfbox{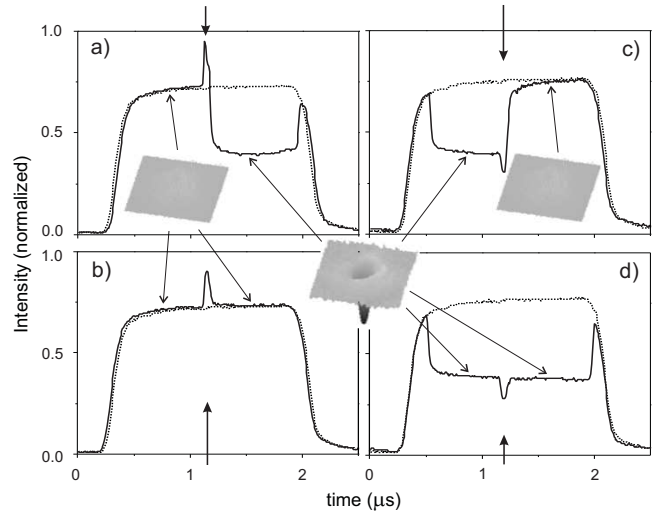}} \vspace{0.7cm}
\caption{Recording of switching-on (a)) and switching-off (c)) of
a bright soliton. "Failed switches" with the address beam
intensity too low to switch soliton on (b)) or off (d)). Heavy
arrows mark the application of address pulses. Dotted traces:
incident intensity. The insets show intensity snapshots, namely
soliton (as Fig.~2a) and unswitched state.}
\end{figure}

\begin{figure}[htbf]
\epsfxsize=60mm \centerline{\epsfbox{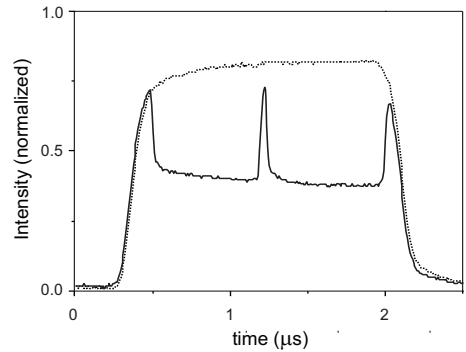}} \vspace{0.7cm}
\caption{Demonstration of the non-thermal character of soliton
switching. As opposed to the incoherent switching [4], the soliton
is stable against local increase of intensity. Dotted trace:
incident intensity.}
\end{figure}

In Fig.~4 we demonstrate that the switching here is really
primarily controlled by the light field in the resonator varying
the carrier density rather than by temperature increase or
generating of phonons as in the incoherent case \cite{tag:4}.

In \cite{tag:4} a soliton could be switched off by increasing
locally the light intensity. The mechanism being that the
additional carrier generation increased the lattice temperature,
which shifts the band edge. With this the existence range of
solitons shifted. The soliton therefore switched off and remained
off until the lattice had cooled again. The address pulse was
perpendicularly polarized to the background field, thus
incoherent with it.

In Fig.~4 the intensity is also increased, but in a coherent way,
by constructive interference with the (in-phase) address pulse.
As opposed to \cite{tag:4} the intensity increase does not switch
the soliton off, showing that a thermal shift of the existence
range does not noticeably occur. Rather here the switching is
controlled through the (fast) electronic nonlinearity.

This absence or negligibility of thermal effects in the formation
of solitons is equally evidenced by the fast spontaneous
switchings in Fig.~3. This nonthermal switching is in agreement
with the experiments \cite{tag:10} where the working wavelength
of the structure used was (equal to the structure grown for there
present observations) slightly below the band edge.

The absence of the thermal effects in the soliton formation was
related in \cite{tag:10} to the predominantly absorptive
character of the nonlinearity. Here essentially the same
conditions prevail. From the working wavelength of 850 nm we
would conclude that the nonlinearity here is even more purely
absorptive than in \cite{tag:10} as the dispersive nonlinearity
should vanish at the precise band edge wavelength. Then the
solitons observed here should rely solely on the absorptive
nonlinearity and the otherwise important mechanism of nonlinear
resonance \cite{tag:11} would be absent.

\begin{figure}[htbf] \epsfxsize=60mm
\centerline{\epsfbox{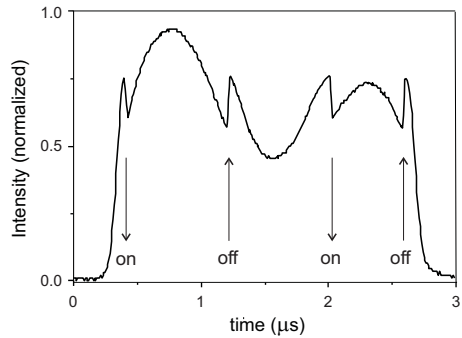}} \vspace{1.0cm} \caption{Repeated
switching on and off of a soliton (see text).}
\end{figure}

Fig.~5 shows fast repeated switching on and off of a soliton. The
phase of the address beam was modulated rapidly here by an
additional electro-optic phase modulator in the address path at a
frequency of ~700 kHz and with a phase excursion of somewhat
larger than $\pi$. Fig.~5 proves again that no slow thermal
process is involved in soliton formation or its switch off.\\

We conclude by noting that these experiments demonstrating the
switching of solitons constitute another step towards application
of semiconductor solitons to optical information processing e.g.
in certain telecom tasks.\\

Acknowledgments\\ This work was supported by the European ESPRIT
Project "PIANOS" and by Deutsche Forschungsgemeinschaft under
grant We743/12-1. Help from C.O.Weiss and helpful discussions
with R.Kuszelewicz are acknowledged.

\end{document}